\documentclass{emulateapj}
\usepackage{graphicx}

\def\be{\begin{equation}}
\def\ee{\end{equation}}
\def\ba{\begin{eqnarray}}
\def\ea{\end{eqnarray}}
\def\go{\mathrel{\raise.3ex\hbox{$>$}\mkern-14mu
             \lower0.6ex\hbox{$\sim$}}}
\def\lo{\mathrel{\raise.3ex\hbox{$<$}\mkern-14mu
             \lower0.6ex\hbox{$\sim$}}}
\def\etal{et al.\ \rm}

\def\bn{{\bf n}}
\def\bm{{\bf m}}
\def\bN{{\bf N}}
\def\bv{{\bf v}}
\def\bs{{\bf s}}
\def\br{{\bf r}}
\def\bR{{\bf R}}
\def\bbeta{\mbox{\boldmath{$\beta$}}}
\def\mus{{\mu\mbox{s}}}

\begin{document}

\title{Effects of Pulsar Rotation on Timing Measurements of the Double
Pulsar System J0737-3039}
\author{Roman R. Rafikov\altaffilmark{1} and Dong Lai\altaffilmark{2}}
\altaffiltext{1}{IAS, Einstein Dr., Princeton, NJ 08540; rrr@ias.edu}
\altaffiltext{2}{Department of Astronomy, Cornell University, 
Ithaca, NY 14853; dong@astro.cornell.edu}


\begin{abstract}
We study the effect of pulsar rotation on timing 
of binary pulsars, with particular emphasis on the 
double pulsar system J0737-3039. Special relativistic 
aberration due to the orbital motion of pulsar 
changes both the longitude and latitude of the 
emission direction with respect to the pulsar spin axis.
The former gives rise to a shift of the arrival time of the pulse
centroid (this is the conventional ``longitudinal'' 
aberration delay), the latter results in a distortion 
(contraction or dilation) of the pulse profile on the orbital timescale. 
In the framework of the rotating vector model of pulsar emission,
the amplitude of pulse distortion depends inversely on 
the variation of polarization position angle across the pulse. 
For small angle between the pulsar magnetic and spin axes, as inferred 
for PSR J0737-3039A from polarimetric observations, 
the pulse distortion is significant ($\sim 1\%$) and the associated 
``latitudinal'' aberration delay is much larger than the longitudinal one.
We show that by monitoring the arrival time of separate 
pulse components as a function of pulsar orbital phase, the 
latitudinal aberration delay may be easily measured with 
the current timing precision. Such measurement would 
constrain the spin geometry of the system.
The latitudinal delay can also be detected by monitoring system's 
orbital parameters on the geodetic precession timescale. 
Because of the near edge-on orbital orientation of the PSR J0737-3039
system, general relativistic bending of pulsar A's radio 
beam near its superior conjunction also introduces spin-dependent
time delays of similar order of magnitude as the aberration delays.
In addition, light bending splits the pulse profile into two variable 
components, corresponding to two gravitationally lensed images of the source.
Detection of lensing effects is challenging, but may be
possible with existing technology.
\end{abstract}


\newpage


\section{Introduction}
\label{sect:intro}


The recently discovered double pulsar system J0737-3039 (Burgay \etal 2003; 
Lyne \etal 2004) presents an unprecedented natural laboratory for 
testing our understanding of general relativity and 
pulsar magnetosphere physics. One of the unique features of this 
system is the almost edge-on orientation of its orbital plane with 
respect to our line of sight. Currently the two best constraints 
on the system's inclination come from the Shapiro delay 
($i=88.7^\circ\pm 0.9^\circ$; Lyne et al.~2004; Ransom \etal 2004) 
and from scintillation measurements 
($|i-90^\circ|=0.29^\circ\pm 0.14^\circ$, Coles \etal 2004). 
Among other effects this high inclination leads 
to periodic eclipses of the millisecond pulsar (pulsar A) by plasma in 
the magnetosphere of the normal pulsar (pulsar B) (see Lyne et al.~2004;
Kaspi \etal 2004; McLaughlin et al.~2004; Arons \etal 2004; 
Rafikov \& Goldreich 2004; Lyutikov \& Thompson 2005). 

The edge-on orientation of the system also makes 
possible gravitational light bending of the pulsar signal 
when the pulsar passes through its superior conjunction with respect to 
the companion. In Lai \& Rafikov (2005), we studied the effect of 
gravitational lensing on the pulse intensity and the impact of 
light bending on the correlated scintillation measurements of both pulsars.
We also evaluated the effects of light bending on the pulse
arrival time: geometric time delay and the modified Shapiro delay
--- both of these timing contributions are independent of the pulsar 
spin.

Since pulsar signals are due to the beamed emission of 
a rotating neutron star (as opposed to radial pulsation of the star), 
additional spin-dependent time delays arise due to special
relativistic aberration (Smarr \& Blandford 1976; Damour \& Deruelle 1986)
and general relativistic light bending (cf.~Schneider 1990).
The goal of this paper is to evaluate these spin-dependent delays
in the PSR J0737-3039 system, and assess their detectability and 
potential for constraining the parameters of the system.
We also examine the distortion of pulse profile due to aberration and
lensing, and discuss an alternative way of analyzing pulse arrival time
and introduce the concept of ``latitudinal delay''
that focuses on specific features in the pulse profile.
For PSR J0737-3039, a measurement of this 
latitudinal delay may lead to useful constraints on the geometry of the
system. 

In \S \ref{sect:timing}, we calculate the spin-dependent time delays 
due to aberration and lensing. In \S 3, we examine the pulse shape
distortion/variation and calculate the latitudinal delays.
The prospects of detecting these effects are discussed in \S 4.

Although the formulae derived in this paper are general for 
any binary pulsar system, we shall apply them to PSR J0737-3039, 
concentrating on the effects of rotational delays 
on the timing of the millisecond pulsar A, with pulsar B playing the 
role of the lensing companion. We adopt the following parameters for the
system: $M_p=M_A=1.337M_\odot,~M_c=M_B=1.25M_\odot$, 
the spin period of pulsar A $P_A=22.7$~ms,
the orbital period $P_b=0.1023~\mbox{d}$,
the orbital semimajor axis $a=8.784\times 10^{10}~\mbox{cm}$,
eccentricity $e=0.0878$, longitude of periastron
$\omega=73.8^\circ$ (as of 2004), and the orbital inclination angle
$i$ in the range between $90.14^\circ$ and $90.56^\circ$.


\section{Spin-Dependent Time Delays}
\label{sect:timing}


Because of special relativistic light aberration, the direction
of emission in the comoving frame of the pulsar, ${\bf N}$ ($|{\bf N}|=1$), 
differs from that in the observer's frame, ${\bf n}$ ($|{\bf n}|=1$), by  
the vector
\be
(\Delta\bn)_A=\bn\times (\bn\times\bbeta_p)
\simeq \bn_0\times (\bn_0\times\bbeta_p),
\label{eq:deln_A}
\ee
where $\bv_p=c\bbeta_p$ is the pulsar velocity relative to the binary
barycenter\footnote{The constant velocity of the barycenter itself does 
not affect timing measurement.}, and ${\bf n}_0$ is the unit vector 
from the binary system to the observer.
Because of general relativistic light bending, the direction 
of emission ${\bf n}$ at the pulsar position differs from 
${\bf n}_0$. 
This lensing effect is important only around the superior 
conjunction of the pulsar, giving rise to potentially measurable 
time delay provided that the minimum projected separation between the
pulsar and its companion is comparable to the companion's
Einstein radius (Lai \& Rafikov 2005). The deflection vector due 
to light bending $(\Delta\bn)_L=\bn-\bn_0$ is given by\footnote{
Analogous to the classical work of Damour \& Taylor (1992),
our calculations of the lensing effects are performed in the static
limit, which neglects the change in the position of the companion
during the time it takes the photon to cross the binary. Including 
this propagation effect in the spirit of Kopeikin \& Sch\"{a}fer (1999) 
would not qualitatively change our conclusions and we omit it from the 
present consideration for the purposes of clarity.}
\be
(\Delta\bn)_L=\left({\Delta R_\pm\over R}\right)
{\bR\over a_\parallel}=
\left({\Delta R_\pm\over R}\right) {\bn_0\times (\br\times \bn_0)\over a_\parallel}.
\label{eq:deln_L}
\ee
Here $\br=\br_p-\br_c$ is the position vector of the pulsar relative to its
companion, $\bR=\bn_0\times (\br\times \bn_0)$ is the projection of $\br$ 
in the sky plane, $R=|\bR|=r(1-\sin^2 i\,\sin^2\psi)^{1/2}$
($\psi$ is the true anomaly measured from the ascending node of the pulsar),
$r=a(1-e^2)/(1+e\cos\phi)$ is the distance between pulsar and its companion
($\phi=\psi-\omega$ is the orbital true anomaly measured from periastron), and 
$a_\parallel=a\sin i (1-e^2)/(1+e\sin\omega)$ is this distance at the 
conjunction projected along our line of sight ($a,e$ and $\omega$ are 
the orbital semimajor axis, 
eccentricity, and longitude of periastron, respectively). Gravitational
lensing gives rise to two pulsar images (specified by the subscript
``$\pm$''); $\Delta R_\pm$ is the displacement (in the sky plane) 
of the image position,
$\bR_\pm=(\Delta R_\pm/R +1)\bR$, relative to the fiducial position,
$\bR$, and is given by 
\be
\Delta R_\pm =
\frac{1}{2}
\left(\pm\sqrt{R^2+4R_E^2}-R\right),
\label{eq:eps}
\ee
with the Einstein radius
\be
R_E=(2R_g a_\parallel)^{1/2}\simeq(2R_g a)^{1/2} =2550~{\rm km}.
\label{eq:ein_rad}
\ee
Here $R_g=2GM_c/c^2=3.69$~km and the numerical value is for J0737-3039A 
with pulsar B 
playing role of the companion.
Combining eqs.~(\ref{eq:deln_A}) and (\ref{eq:deln_L}), the direction of 
emission in the pulsar's comoving frame is $\bN=\bn_0+\Delta\bN$, with 
$\Delta\bN=(\Delta\bn)_A+(\Delta\bn)_L$. 
The deflection angle of the emission vector 
due to light bending, $\Delta R/a$, has a maximum value 
(for the ``plus'' image) $R_E/a\sim v/c$ (this occurs 
at the orbital conjunction and $i=90^\circ$), which is 
of the same order as the aberration effect. 
This similarity 
of the magnitudes of two effects holds 
only in the binary systems which are so highly inclined 
with respect to our line of sight that $R \sim R_E$ 
at conjunction (in more face-on systems with $\cos i\sim 1$
the gravitational light deflection is a $(v/c)^2$ effect).

Variation of the emission direction ${\bf N}$ causes a shift in the 
equatorial longitude $\Phi$ of ${\bf N}$ in the corotating frame of 
pulsar (counted in the direction of rotation) given by
\be
\Delta\Phi=\frac{\Delta{\bf N}\cdot ({\bf s}_p\times{\bf n}_0)}
{|{\bf s}_p\times{\bf n}_0|^2},
\label{eq:delta_phi}
\ee
where $\bs_p$ is the unit vector along the pulsar spin axis.
For an arbitrary emission pattern rigidly rotating 
around ${\bf s}_p$, any change in the equatorial longitude corresponds
to a change of emission phase, leading to a time delay 
$\Delta t=\Delta\Phi/\Omega_p=(\Delta t)_A+(\Delta t)_L$
(here $\Omega_p$ is the angular frequency
of the pulsar and the time delay is positive for signal arriving later),
where
\begin{eqnarray}
&& (\Delta t)_A=-\frac{\bbeta_p\cdot (\bs_p\times{\bf n}_0)}
{\Omega_p\,|{\bf s}_p\times{\bf n}_0|^2},
\label{eq:aber_delay}\\ 
&& (\Delta t)_L=
\left({\Delta R_\pm\over R}\right){ \br\cdot (\bs_p\times\bn_0)\over
\Omega_p a_\parallel |\bs_p\times\bn_0|^2}
\label{eq:lensing_delay}
\end{eqnarray}
are contributions due to aberration (Smarr \& Blandford 1975) 
and gravitational light bending (cf. Schneider 1990)
\footnote{This agrees with with expression 
of Schneider (1990) if we replace $a_\parallel$ by 
$\br\cdot \bn_0=r\sin i\sin\psi$ --- this would lead to divergence at 
$\psi=0,\pi$, which is naturally avoided in our case.},
respectively. 
Note that this ``longitudinal'' time delay shifts the whole 
pulse profile uniformly without introducing any distortions to its 
shape (see \S 3). 

The pulsar spin axis $\bs_p$ is specified by two angles 
(see Fig.~1 of Damour \& Taylor 1992; thereafter DT92): $\zeta$ is 
the angle between $\bs_p$ and the light-of-sight vector $\bn_0$,
and $\eta$ is the angle between the ascending node of the orbit and the 
projection of $\bs_p$ on the sky plane.  The aberration time delay 
(\ref{eq:aber_delay}) is then given by
\be
(\Delta t)_A=A \left(\sin\psi+e\sin\omega\right)+
B\left(\cos\psi+e\cos\omega\right),
\label{eq:aber_delay1}\ee
with
\begin{eqnarray}
\left\{\begin{array}{l}A\\ B
\end{array}\right\}=-\frac{\Omega_b a_p}{\Omega_p c\sqrt{1-e^2}\sin\zeta}
\left\{\begin{array}{l}
\sin\eta\\ \cos i\cos\eta
\end{array}\right\},
\label{eq:AB_A}\end{eqnarray}
where $\Omega_b=(GM_t/a^3)^{1/2}$ ($M_t=M_p+M_c$ is the total 
mass of the system) and $a_p=(M_c/M_t)a$. 
These expressions coincide with the results of Damour \& Deruelle (1986)
and DT92 (expressed in terms of a complimentary angle 
$\lambda=\pi-\zeta$ instead of $\zeta$). 
Similarly, the delay associated with light bending is
\begin{eqnarray}
(\Delta t)_L &=& -\left({\Delta R_\pm\over R}\right)
\!\!\left({r\over a_\parallel}\right)
{\sin\eta\,\cos\psi-\cos i\,\cos\eta\,\sin\psi \over \Omega_p \sin\zeta}
\nonumber\\
&=& -\left({\Delta R_\pm\over a_\parallel\Omega_p}\right)
{(\sin\eta\,\cos\psi-\cos i\,\cos\eta\,\sin\psi)\over
\sin\zeta\,(1-\sin^2i\,\sin^2\psi)^{1/2}}.
\label{eq:dtl}
\end{eqnarray}
Note that the lensing effect is important only around the 
orbital conjunction ($\psi=\pi/2$) for nearly edge-on systems.
Let $i=\pi/2+\Delta i$ and $\psi=\pi/2+\Delta\psi$, with $|\Delta i|,
|\Delta\psi|\ll 1$, then we have
\be
(\Delta t)_L\simeq -\left({\Delta R_\pm\over a_\parallel\Omega_p}\right)
{\Delta i\, \cos\eta-\Delta\psi\,\sin\eta\over
\sin\zeta \left[(\Delta i)^2+(\Delta\psi)^2\right]^{1/2}},
\label{eq:deltatl}\ee
with $\Delta R_\pm$ given by eq.~(\ref{eq:eps}) and 
$R\simeq a_\parallel \left[(\Delta i)^2+(\Delta\psi)^2\right]^{1/2}$.
Both eqs.~(\ref{eq:lensing_delay}) and (\ref{eq:dtl}) differ 
from the result of Doroshenko \& Kopeikin (1995) 
(also quoted in Wex \& Kopeikin 1999)
since these authors used the equation for the light ray trajectory
from Klioner \& Kopeikin (1992) obtained under the assumption of
$R\gg R_E$. As a result, in the most 
interesting case when $R\sim R_E$ (and bending delay attains its 
maximum value) the expression for $(\Delta t)_L$ obtained by Doroshenko 
\& Kopeikin (1995) 
is not applicable and cannot be used for interpreting timing measurements 
\footnote{By contrast, our expression is valid for all $R$ as long as
$R\ll a_\parallel$. For 
$R\sim a_\parallel$, 
the lensing effect is
completely negligible.}. Also note that some previous 
treatments of bending delay (e.g. Goicoechea \etal 1995)
were restricted to circular and/or purely edge-on orbits.

Neglecting the dependence on various angles, the aberration time delay 
is of order $\Omega_b a_p/(c\Omega_p)=3.6\,\mu s$,
and the rotational lensing delay at the conjunction is of order
$\Delta R_\pm /(a_\parallel\Omega_p)=10.5 (\Delta R_\pm/R_E)\,\mu s$.
Not surprisingly, the two delays are of the same order of magnitude 
at the conjunction, since $\Omega_b a_p\sim v_p/c\sim v/(2c)$ and
$R_E/a\sim \sqrt{2} v/c$.

\begin{figure}
\plotone{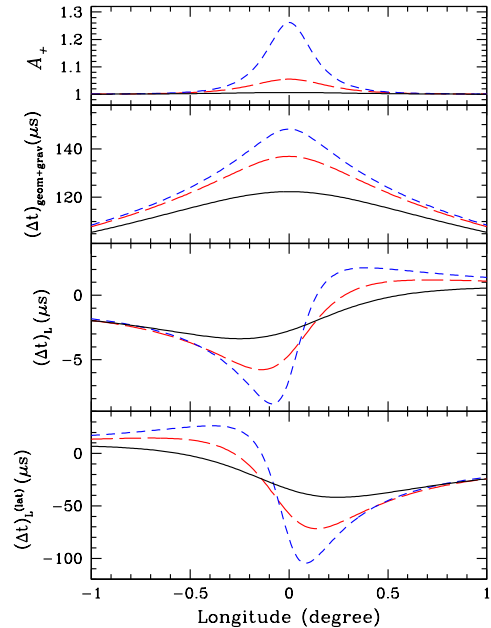}
\caption{The amplification (top panel), the combined geometric and
gravitational (Shapiro) delay (the second panel), the rotational
lensing delay (the third panel), and the latitudinal lensing
delay (the bottom panel) of the dominant (``plus'')
image of the pulsar A signal as a function of the orbital phase.
The longitude is measured from the superior conjunction of pulsar A 
(when A is exactly behind pulsar B).
In each panel, the inclination angles are $i=90.56^\circ$ (solid line),
$90.28^\circ$ (long-dashed line) and $90.14^\circ$ (short-dashed line).
For the third and bottom panels, the angles $\zeta=50^\circ$, 
$\eta=45^\circ$ and $\tan\chi_0=0.08$ are used.
\label{fig:pos_image}}
\end{figure}

\begin{figure}
\plotone{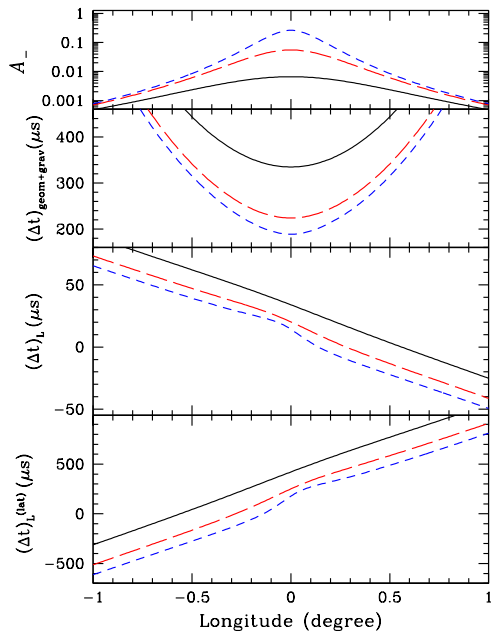}
\caption{Same as Figure \ref{fig:pos_image}, except for the 
subdominant (``minus'') image. 
\label{fig:neg_image}}
\end{figure}

Damour and Deruelle (1986) recognized that the aberration delay
is degenerate with the Roemer delay, and therefore cannot be directly 
measured (but see \S 4). The time delays associated with lensing do not
suffer from this degeneracy. Figures \ref{fig:pos_image} and 
\ref{fig:neg_image} show these delays
(the combined geometric and Shapiro delays as calculated in 
Lai \& Rafikov (2005), and the rotational lensing delay
given by eq.~[\ref{eq:dtl}] or [\ref{eq:deltatl}]) together with the
amplification factor for both images,
\be
{\cal A}_\pm={u^2+2\over 2u\sqrt{u^2+4}}\pm {1\over 2},\qquad u=R/R_E.
\label{eq:mag}\ee
In evaluating $(\Delta t)_L$, we adopt $\zeta=50^\circ$
as suggested by Demorest et al.~(2004) from polarization measurement
(although this angle is currently unconstrained)
and use $\eta=45^\circ$ as an example in Figures \ref{fig:pos_image} and 
\ref{fig:neg_image}. 

One can see from Figure \ref{fig:pos_image} that 
near conjunction the lensing delay associated with positive image 
reaches $\approx 6~\mu$s for $i=90.28^\circ$. In binary systems with smaller 
inclination angles the timing signal due to lensing is weaker. One
can show that if the minimum projected separation of pulsar from its 
companion $a_\parallel|\Delta i|\gg R_E$ then for $e\ll 1$ the maximum amplitude of
$(\Delta t)_L$ is 
\be
{\rm max}|(\Delta t)_L|\approx \frac{(R_E/a_\parallel)^2}
{2|\Delta i|\Omega_p\sin\zeta}(1+|\cos\eta|),
\label{eq:max_t_L}
\ee
reached at $\Delta\psi\approx \Delta i(\cos\eta\mp 1)/\sin\eta$, where the upper
sign corresponds to $\cos\eta>0$, while the lower one to $\cos\eta<0$. Note that 
increasing $|\Delta i|$ diminishes ${\rm max}|(\Delta t)_L|$ no faster than  
$|\Delta i|^{-1}$. The orbital phase at which this maximum occurs shifts 
from conjunction in proportion to $|\Delta i|$ reducing 
magnetospheric absorption effects in systems with larger $|\Delta i|$.
In particular, if the real inclination angle of J0737-3039 is closer to that
measured with the Shapiro delay ($i=88.7^\circ$) rather than the one inferred 
using the scintillation technique, then ${\rm max}|(\Delta t)_L|$ amounts to 
$\approx 1.7~\mu$s (which is at present unmeasurable) reached $\approx 13$ s
after conjunction, which is almost at the end of eclipse of pulsar A.

The magnitude of $(\Delta t)_L$ scales as $(\sin\zeta)^{-1}$
and its shape depends on $\eta$ as $(\Delta i)\cos\eta-(\Delta\psi)\sin\eta$.
As a result, unlike the geometric delay and Shapiro delay, the rotational 
lensing delay is generally {\it asymmetric} with respect to orbital 
conjunction (the exception is when $\eta=0,~180^\circ$).
Note that although the delays associated with the ``minus'' image
can become very large ($100$'s of $\mu s$) as a result of its large
deflection ($|\Delta R_-/a_\parallel|$), 
this image is usually too demagnified to be of interest
(see \S \ref{subsec:double}).

\section{Pulse Profile Variation and ``Latitudinal Time Delay''}

\subsection{Pulse Profile Variation}
\label{subsect:pulse}

Variation of the pulse emission direction ${\bf N}$ also results in 
the change of colatitude of emission vector, $\zeta$. Since 
$\cos\zeta=\bs_p\cdot\bN$ and $\sin\zeta=|\bs_p\times\bN|$, we find
\be
\Delta\zeta=-\frac{{\bf s}_p\cdot \Delta{\bf N}}
{|{\bf s}_p\times{\bf n}_0|}=(\Delta\zeta)_A +
(\Delta\zeta)_L,
\label{eq:delta_lambda}\ee
where
\begin{eqnarray}
&& (\Delta\zeta)_A = -(\bn_0\times\bbeta_p)\cdot 
{(\bs_p\times\bn_0)\over |\bs_p\times\bn_0|}\nonumber\\
&& = \frac{\Omega_b a_p}{c\sqrt{1-e^2}}\left[
\cos i\sin\eta\left(\cos\psi+e\cos\omega\right)\right.
\nonumber\\
&& ~~~~~~~~~~~~~~~~\left.-
\cos\eta\left(\sin\psi+e\sin\omega\right)\right],
\label{eq:del_lam_A}\\
&& (\Delta\zeta)_L = {\Delta R_\pm (\bn_0\times \br)
\over a_\parallel R}\cdot{(\bs_p\times\bn_0)\over |\bs_p
\times\bn_0|}\nonumber\\
&& =-
\left({\Delta R_\pm\over R}\right) {r\over a_\parallel}
(\cos\eta\cos\psi+\cos i\sin\eta\sin\psi)
\label{eq:delta_lambda_L}
\end{eqnarray}
are the contributions from aberration and lensing, respectively.

\begin{figure}
\plotone{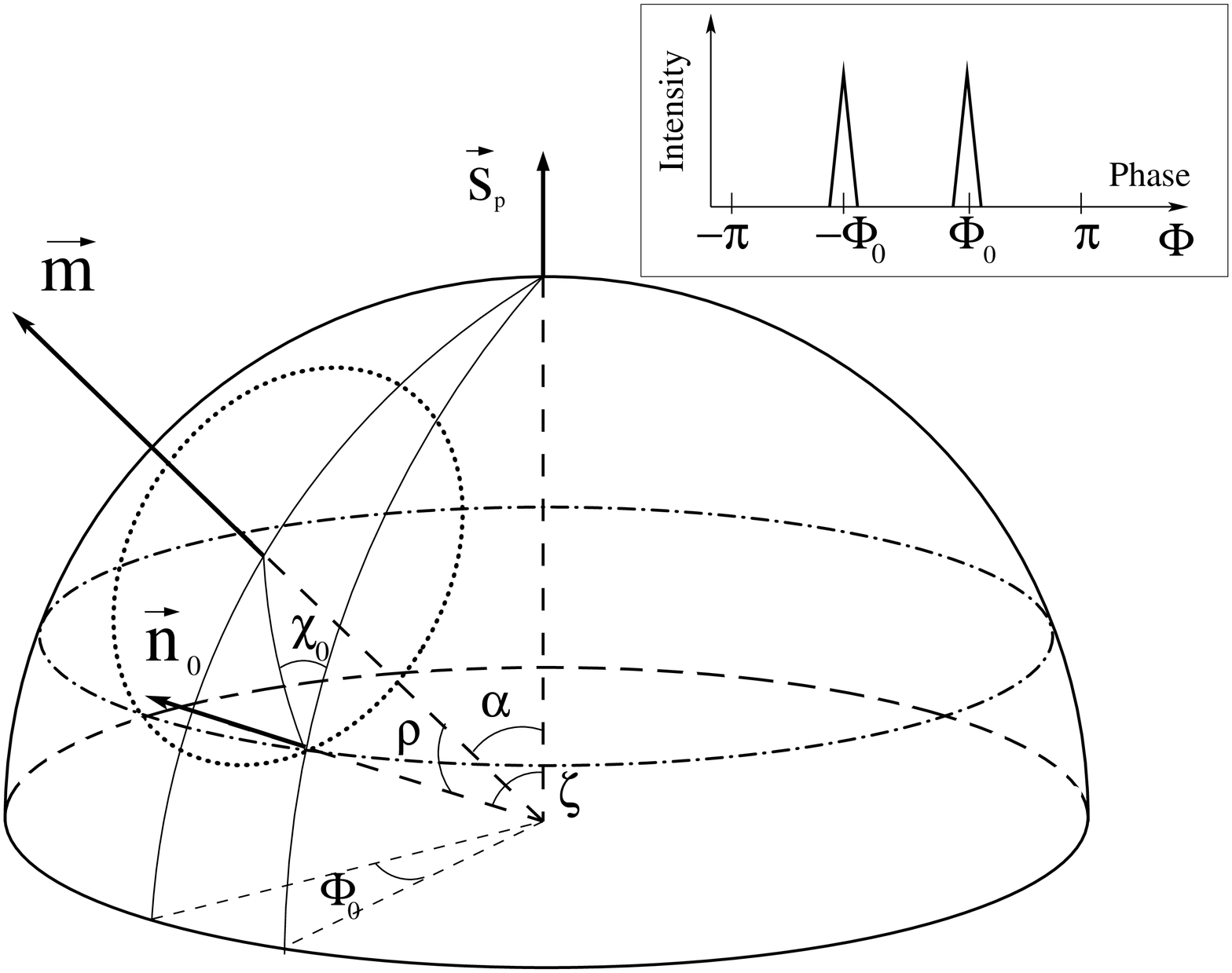}
\caption{
Geometry of pulsar emission with respect to the spin
axis ${\bf s}_p$, magnetic axis ${\bf m}$, and our line 
of sight ${\bf n}_0$. The latter cuts the celestial sphere and
emission pattern (cone with opening angle $\rho$; the dotted 
circle denotes its intersection with celestial sphere) along 
the dot-dashed line in the course of pulsar rotation, giving rise to 
features of the pulse profile at $\pm\Phi_0$, shown schematically 
in the inset.
\label{fig:mag_pole}}
\end{figure}

Unlike the longitudinal shift of the emission vector, 
the shift of colatitude affects the different 
components of pulse profile differently, thus changing the pulse 
profile periodically on the orbital timescale.  
In the following we consider for illustrative purposes a specific 
radio emission pattern
based on the rotating vector model (RVM; Radhakrishnan \& Cooke 1969) 
widely used for interpreting pulsar pulse profile and polarization data.
In this model, the radio emission pattern consists of a set of cones which 
are circular in cross section as seen along the magnetic axis
${\bf m}$ (see Fig.~\ref{fig:mag_pole}).
As a result of pulsar rotation around $\bs_p$, our line of sight 
${\bf n}_0$ passes through the emission cone with half opening angle 
$\rho>|\zeta-\alpha|$ (where $\alpha$ is the angle between $\bs_p$ 
and $\bm$) which causes two episodes of radio emission at 
the pulse phases $\pm \Phi_0$ corresponding to the leading 
and trailing edges of the cone\footnote{The radio intensities
of the two emission episodes are generally different because of the
complicated azimuthal structure of the emitting region.}.
Emission patterns close to circular are supported by theoretical
ideas of pulsar emission mechanisms and observations of slow 
pulsars (Rankin 1993), although their occurrence in millisecond pulsars 
is not so clear (Kramer \etal 1998; Weisberg \& Taylor 2002). 
The observed pulse profile of PSR J0737-3039A is consistent with
a circular emission pattern (Demorest et al.~2004; Manchester et al.~2005).

\begin{figure}
\plotone{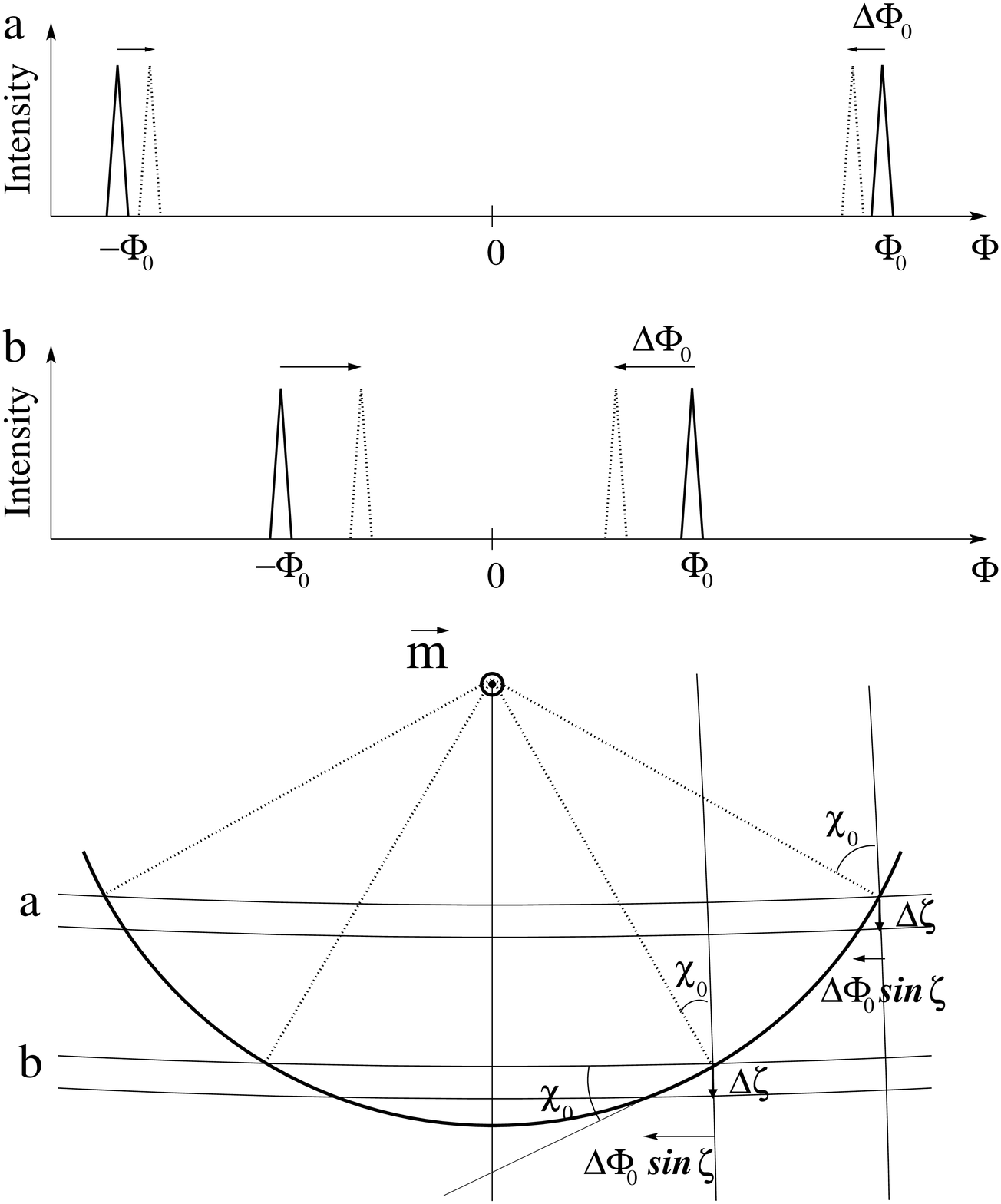}
\caption{
Variation of the pulse profile as a result of colatitudinal variation 
$\Delta \zeta$ for two different cuts of our 
line of sight $\bn_0$ through the circular emission pattern.
The bottom panel shows the projection onto the pulsar's celestial sphere 
along the magnetic axis: the thick solid arc depicts part of the emission cone,
the thin vertical arcs represent projections of meridians, 
and the dotted lines denote the magnetic field lines, a and b
denote the locus of two different lines of sight. This Figure illustrates how 
the proximity of our line of sight to the edge 
of the emission cone affects the total pulse width.
\label{fig:projection}}
\end{figure}

The full width of the pulse is $2\Phi_0$, with
\be
\cos\Phi_0={\cos\rho-\cos\zeta\cos\alpha\over\sin\zeta\sin\alpha}.
\ee
The variation of $\zeta$ then leads to a variation of $\Phi_0$:
\be
\Delta\Phi_0=\Delta\zeta\left(
\frac{1}{\tan\Phi_0\tan\zeta}-\frac{1}{\sin\Phi_0\tan\alpha}\right).
\label{eq:ampl1}
\ee
It is easy to show that $\Delta\Phi_0$ is given by 
(which is also clear from Fig.~\ref{fig:projection})
\be
\Delta\Phi_0=-{\Delta\zeta\over \sin\zeta\tan\chi_0},
\label{eq:delphi0}
\ee
where $\chi_0$ is the angle (on the celestial sphere) between the 
arc connecting $\bn_0$ and $\bs_p$ and the arc connecting $\bn_0$ 
and $\bm$ at the edges of the pulse, i.e., it coincides with the position 
angle of linear polarization and is given by the usual expression
(Komesaroff 1970)
\be
\tan\chi_0=\frac{\sin\alpha\sin\Phi_0}{\cos\alpha\sin\zeta-
\cos\Phi_0\sin\alpha\cos\zeta}.
\label{eq:chi}
\ee
Equation (\ref{eq:delphi0}) shows that in the framework of RVM 
the change of pulse width 
depends inversely on the change of position angle of radio polarization 
across the pulse. It also shows that while $\Delta\zeta$ is
generally quite small ($\sim v_p/c\sim 10^{-3}$ due to aberration
and $\sim \Delta R_\pm/a\sim 3\times 10^{-3}\Delta R_\pm/R_E$ due
to lensing), $\Delta\Phi_0$ can be significant when 
$\sin\zeta\tan\chi_0$ is small, as in the case of the PSR J0737-3039
system (see \S\ref{subsec:colat}).

Note that the variation of the pulse profile on the orbital 
timescale due to the colatitudinal shift is not a unique property of 
RVM --- this effect should exist for more complicated emission 
patterns as well. Interpretation of a positive detection of this effect
in terms of pulsar spin geometry would be less straightforward 
for intricate beam shapes, and the exact relationship between 
$\Delta\Phi_0$ and $\chi_0$
given by eq.~(\ref{eq:delphi0}) would also break 
down, but they do not affect our basic conclusion about the existence 
of such a phenomenon.

\subsection{Latitudinal Time Delay}
\label{subsec:colat}

Standard pulsar timing formulae (including those discussed
in \S 2) measure the arrival time of the {\it centroid} of radio pulses.
However, pulse profile often has multiple components; in the rotating
vector model these components correspond to passages of $\bn_0$ through
different emission cones (two for each cone). As discussed above, the 
variation of colatitude of
emission direction due to aberration and lensing results in 
the change of the location (in phase) of a given pulse component relative to
the pulse centroid. Thus if one times the arrival
of a specific component of the pulse, additional delay associated with 
such shift of colatitude must be included; we call this ``latitudinal
delay''.

In RVM, the latitudinal delay associated with the
leading edge of the pulse is\footnote{For the trailing edge, the delay is
$-\Delta\Phi_0/\Omega_p$. Hereafter our formulae refer to the leading edge
of the pulse.}
\be
(\Delta t)^{(\rm lat)}=\Delta\Phi_0/\Omega_p=
(\Delta t)_A^{(\rm lat)}+(\Delta t)_L^{(\rm lat)},
\ee
where $(\Delta t)_A^{(\rm lat)}$ and $(\Delta t)_L^{(\rm lat)}$ are
contributions due to aberration and lensing.
Using eqs.~(\ref{eq:del_lam_A}) and (\ref{eq:delphi0}), we find
the latitudinal aberration delay 
\begin{eqnarray}
(\Delta t)_A^{\rm (lat)}=C\left(\sin\psi+e\sin\omega\right)+
D\left(\cos\psi+e\cos\omega\right),
\label{eq:d_t_c_A}
\end{eqnarray}
where
\begin{eqnarray}
\left\{
\begin{array}{l}
C\\
D
\end{array}
\right\}=-\frac{\Omega_b a_p}{\Omega_p c\sqrt{1-e^2}}
\frac{1}{\sin\zeta\tan\chi_0}
\left\{
\begin{array}{l}
-\cos\eta\\
\cos i\sin\eta
\end{array}
\right\},
\label{eq:C_and_D}
\end{eqnarray}
with $\tan\chi_0(\Phi_0)$ given by (\ref{eq:chi}) in the framework of RVM.
Similarly, using eq.~(\ref{eq:delta_lambda_L}), we find the
latitudinal lensing delay
\begin{eqnarray}
&& (\Delta t)_L^{\rm (lat)} = \left({\Delta R_\pm\over R}\right) 
{r\over a_\parallel}{\cos\eta\cos\psi+\cos i\sin\eta\sin\psi\over
\Omega_p\sin\zeta\tan\chi_0}
\nonumber\\
&& ~~\simeq -\left({\Delta R_\pm\over a_\parallel\Omega_p}\right)
{\Delta i\,\sin\eta+\Delta\psi\,\cos\eta\over
\sin\zeta\,\tan\chi_0\,[(\Delta i)^2+(\Delta\psi)^2]^{1/2}},
\label{eq:delt_L_col}
\end{eqnarray} 
where the second equality comes from expansion around
$\psi=\pi/2+\Delta\psi$ and $i=\pi/2+\Delta i$.

Naturally, if one times the arrival of a specific component of
pulse profile, then the total delay is the sum of
the longitudinal contributions [eqs.~(\ref{eq:aber_delay1}),~(\ref{eq:dtl})]
associated with the shift of the pulse centroid and the latitudinal 
contributions [eqs.~(\ref{eq:d_t_c_A}),~(\ref{eq:delt_L_col})] associated 
with pulse dilation/shrinkage.  Thus,
the combination of longitudinal and latitudinal delays results 
in (1) a uniform shift of the whole pulse profile in time by 
$(\Delta t)_A+(\Delta t)_L$ and (2) a nonuniform profile 
contraction/dilation, with each pulse component 
being shifted in phase by $\Delta \Phi_0=
\Omega_p[(\Delta t)_A^{(\rm lat)}+(\Delta t)_L^{(\rm lat)}]$. Pulse 
components closer to the pulse centroid (i.e. those with smaller $\Phi_0$)
are relatively more susceptible to contraction/dilation than the more 
distant components.

To evaluate $(\Delta t)^{(\rm lat)}$, a knowledge of $\zeta,~\chi_0$
and $\eta$ is necessary. Analysis of polarization data of 
PSR J0737-3039A constrains the angle between the spin axis and
the magnetic axis to be $\alpha\approx 4^\circ$, while leaving
$\zeta$ unconstrained (Demorest et al.~2004). The pulse profile of pulsar A 
exhibits remarkable similarity between its outermost leading 
and trailing edges (Demorest \etal 2004; Manchester et al.~2005) at 
$\Phi_0\approx 115^\circ$, which suggests that all profile components 
come from the same magnetic pole and that the two prominent spikes in the 
pulse profile can be associated with leading and 
trailing edges of the same emission cone. Based on the
opening angle of other millisecond pulsars and $\zeta\sim\rho$,
Demorest et al.~(2004) suggests $\zeta\sim 50^\circ$, which we adopt 
in our estimate. 
Equation (\ref{eq:chi}) then yields $\tan\chi_0\approx 0.08$. 
Such 
small value of $\tan \chi_0$ results from the fact
that in the Demorest et al. (2005) geometry our line of sight
passes through the emission cone of pulsar A very close 
to its edge, see their Figure 3 for illustration.

The bottom panels of Figures \ref{fig:pos_image} and \ref{fig:neg_image} 
depict the lensing latitudinal
delay $(\Delta t)_L^{\rm (lat)}$ for both images.
Clearly, because of the small value of $\tan\chi_0$ for
the PSR J0737 system, $(\Delta t)_L^{\rm (lat)}$ is much larger
than $(\Delta t)_L$
and reaches $\approx 79~\mu$s for $i=90.28^\circ$. In binary systems 
with smaller inclination angles satisfying $a_\parallel|\Delta i|\gg R_E$ 
the maximum amplitude of $(\Delta t)_L$ reached at 
$\Delta\psi\approx \Delta i(1-\sin\eta)/\cos\eta$  is given by
\be
{\rm max}|(\Delta t)_L^{\rm (lat)}|\approx \frac{(R_E/a_\parallel)^2}
{2|\Delta i|\Omega_p\sin\zeta\tan\chi_0}(1+\sin\eta),
\label{eq:max_t_L_lat}
\ee
for $e\ll 1$. Scaling of ${\rm max}|(\Delta t)_L^{\rm (lat)}|$ with 
$|\Delta i|$ is analogous to that of ${\rm max}|(\Delta t)_L|$. 
For $i=88.7^\circ$ and spin-magnetic orientation of pulsar A 
adopted here one finds ${\rm max}|(\Delta t)_L^{\rm (lat)}|
\approx 22~\mu$s (which is likely still measurable) reached $\approx 13$ s
before conjunction, which is at the ingres of eclipse of pulsar A.

Another constraint on the pulsar A spin orientation (independent 
from the polarization measurements) comes from the observed lack of 
profile variation of A which should occur in general because of 
A's geodetic precession (Manchester \etal 2005). Stability of A's pulse
profile suggests that A may currently be at a special precessional phase
with its spin axis almost coplanar with both the orbital
angular momentum axis and our line of sight $\bn_0$, implying $\eta\sim
90^\circ$. Assuming this Manchester \etal (2005) have determined the region of 
$\alpha$-$\zeta$ phase space compatible with A's pulse profile stability. 
A wide range of $\alpha$ is allowed, including the small-$\alpha$
solution preferred by Demorest et al.~(2004). In Table \ref{table} we list 
the values of timing coefficients
$A$, $B$, $C$, $D$, together with the maximum amplitudes 
of longitudinal and latitudinal lensing delays (for $i=90.28^\circ$) 
for several spin orientations
of A suggested by Demorest \etal (2004) and Manchester \etal (2005),
assuming for illustrative purposes that in the latter case $\eta=85^\circ$.
Clearly, for small $\alpha$ (hence small $\tan\chi_0$), the latitudinal 
delays are much larger than the longitudinal ones, although the
latitudinal aberration delay is reduced for $\eta\sim 90^\circ$
(while the magnitude of the latitudinal lensing delay is largely
unaffected). 
For illustrative purposes, 
in the rest of the paper 
we will focus on the small-$\alpha$ solution (Demorest \etal 2004).

\subsection{Effect of Double Images}
\label{subsec:double}

In addition to the pulse profile change due to the change of
emission colatitude (see \S\ref{subsect:pulse}), gravitational
lensing affects pulse profile in another way.
Near the orbital conjunction, light bending splits
the source (pulsar) into two radio images in the plane of the sky.
The displacement of the image from the source is $\Delta R_\pm$, 
as given by eq.~(\ref{eq:eps}), and the image magnification 
${\cal A}_\pm$ given by eq.~(\ref{eq:mag}).
While aberration affects the arrival time of both images in the same way, 
lensing breaks this degeneracy and separates the arrival times 
of two weighted replicas of the unperturbed pulse profile.
Assuming that in the absence of orbital motion and lensing 
the pulse profile is given by $I_0(\Phi)$, we find
that the observed pulse profile at orbital phase $\psi$,
including the light bending effect, is
\begin{eqnarray}
I(\Phi;\psi)=&& {\cal A}_+\, e^{-\tau_+}
I_0\left(\Phi-\Omega_p\Delta t_+\right)\nonumber\\
&& +{\cal A}_-\, e^{-\tau_-}I_0\left(\Phi-\Omega_p\Delta t_-\right),
\label{eq:real_shape}
\end{eqnarray}
where the factors $e^{-\tau_\pm}$ accounts for the absorption of
pulsar signal by plasma in the magnetosphere of companion.
The time delays $\Delta t_\pm$ 
for both images include not only the rotational contributions 
[eqs.~(\ref{eq:aber_delay1}),~(\ref{eq:dtl}),~(\ref{eq:d_t_c_A}) and
(\ref{eq:delt_L_col})], but also the geometric delay and
the (modified) Shapiro delay (see Lai \& Rafikov 2005).
Note that $A_\pm,~\tau_\pm,~\Delta t_\pm$ are all functions of
the orbital phase $\psi$.

\begin{figure}
\plotone{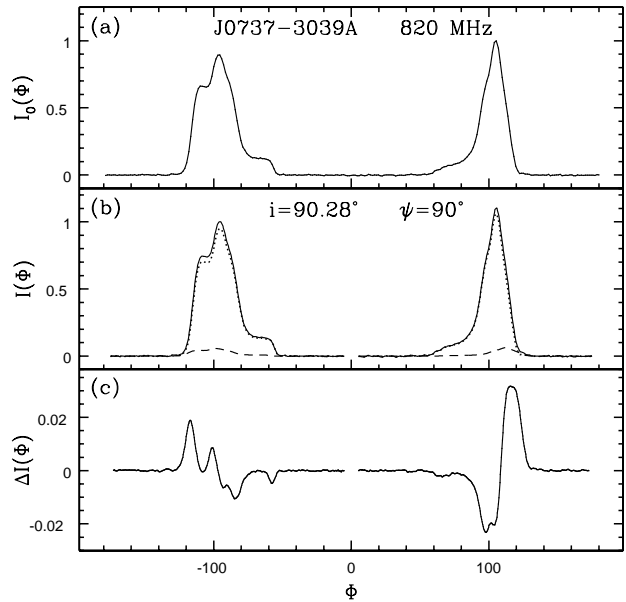}
\caption{
(a) Pulse profile of J0737-3039A far from conjunction, $I_0(\Phi)$. 
(b) Contributions to the total pulse profile $I(\Phi)$ ({\it solid curve})  
at conjunction ($\psi=90^\circ$) due to the positive ({\it dotted curve})
and negative lensed image ({\it dashed curve}) taking into account different
time delays and magnifications of the images. 
In the calculation $i=90.28^\circ$, $\zeta=50^\circ$, $\eta=45^\circ$,
and $\alpha=4^\circ$ are assumed. 
(c) Difference between $I(\Phi)$ and $I_0(\Phi)$,
with the latter magnified by ${\cal A}_++{\cal A}_-$ and shifted by
$\Delta t_+$ in time at conjunction. The remaining residuals are due to 
the improperly subtracted negative image and are rather small. 
The flux units in the three panels are the same.
\label{fig:shape}}
\end{figure}

For illustrative purposes, we assume $\tau_+=\tau_-$, and show in 
Figure \ref{fig:shape} the relative contributions of the two images 
in shaping the pulse profile $I(\Phi)$ near the orbital conjunction for 
$i=90.28^\circ$. In practice, one could adopt a simple
procedure when analyzing arrival times near conjunction: 
since the ``+'' image dominates the 
lightcurve, one would naturally assume that the timing residual 
due to lensing is primarily given by $\Delta t_+$ and the negative image 
contributes to the pulse profile mainly through its magnification. 
Figure \ref{fig:shape}c depicts the relative error incurred 
by this assumption: we plot the difference between $I_0$ magnified by 
${\cal A}_++{\cal A}_-$ and shifted by $\Delta t_+$ in time and the 
real profile $I(\Phi)$ which properly includes contributions from both 
images at conjunction.  One can see that this simple procedure does
perform quite well, and the flux residuals due to improper subtraction of the
negative image are at most several percent. 
In reality, since the light of ``$-$'' image passes closer to the companion 
star than the light of ``+'' image, and the magnetospheric absorption 
depends strongly on the plasma density, we expect that near 
conjunction, $\tau_-$ is much greater than $\tau_+$. This would further 
decrease the contribution of the ``$-$'' image to the distortion of the 
pulse profile.

\section{Detectability of Rotational Delays and Pulse Distortion}
\label{sect:detect}

The rotational time delays discussed in previous sections 
could be observed in different ways. The aberration effect
results in arrival time variation and pulse distortion
throughout the whole cycle of the orbital motion, while the 
lensing effect is important only in very edge-on systems near 
the superior conjunction of the pulsar.
As discussed above, the pulse profile and polarization of 
PSR J0737-3039A imply that our line of sight crosses the edge of the 
pulsar's emission cone (Demorest et al.~2004), resulting in
significant pulse distortion and the associated latitudinal delay.
For example, taking $\zeta=50^\circ$, $\tan\chi_0=0.08$ and 
$i=90.28^\circ$, we find, for the aberration delays, 
$A=-4.7\sin\eta~\mus$, $B=0.023\cos\eta~\mus$, $C=59\cos\eta~\mus$ and 
$D=0.3\sin\eta~\mus$. The lensing delays are shown in Figures 
\ref{fig:pos_image} and \ref{fig:neg_image}.
Aberration results in fractional pulse width variation of order
$\Delta\Phi_0/\Phi_0=\Omega_p(\Delta t)_A^{(\rm lat)}/\Phi_0
\sim \Omega_p C/\Phi_0\sim 10^{-2}\cos\eta$
in the course of the pulsar's orbital motion. For a few seconds
around the superior conjunction, light bending gives rise to 
additional profile distortion at a similar level.

Motivated by our results, we propose to perform timing of 
{\it separate components} of the pulse profile to detect
pulse dilation/contraction and the associated latitudinal delays.
Instead of fitting one template to the whole observed pulse profile 
to determine its overall timing displacement (or the arrival time of the
pulse centroid), one can directly fit templates for each 
individual component of the profile. Assuming that 
some pairs of components represent the 
leading and trailing edges of the emission cone through our line of sight, 
one can measure the difference between their times of arrival 
and correlate its variation with the orbital phase. 
This method presents a relatively easy way of measuring the latitudinal
delays and extracting pulsar parameters from them. 
In the case of PSR J0737-3039A, periodic profile distortions 
must be very strong because of the small ($\sim 4^\circ$)
magnetic inclination angle of pulsar A with respect to its spin
(Demorest \etal 2004). The latitudinal delay is of order
$C=59\cos\eta~\mus$, which should be detectable since 
RMS timing residuals of pulsar A are already at the 
level of $27~\mu$s (Lyne \etal 2004). Such detection would directly
constrain $\chi_0,~\zeta$ and $\eta$.

Detection of the standard ``longitudinal'' aberration delay 
(eqs.~[\ref{eq:aber_delay1}]-[\ref{eq:AB_A}]) is a more difficult task. 
Damour \& Deruelle (1986) were the first to recognize that this delay 
cannot be observed directly as a timing variation on the 
orbital time scale since it can be fully absorbed into the Roemer delay
by rescaling/redefining the eccentricity, semimajor axis, epoch of
periastron and two other post-Newtonian parameters.
More precisely, because of the aberration delay, the observed values of 
the semimajor axis $a^{obs}$, eccentricity $e^{obs}$, and the epoch 
of periastron $T_0^{obs}$ are related to the true values of 
these quantities $a^{true}, e^{true}, T_0^{true}$ by the formulae
\begin{eqnarray} 
&& \frac{a^{obs}}{a^{true}}=\frac{e^{obs}}{e^{true}}=1+\epsilon_A,\nonumber\\
&& \epsilon_A\equiv \frac{c A}{a_p\sin i}\approx -3.3\times 10^{-6}\sin\eta, 
\label{eq:ae_rel}\\
&& T_0^{obs}=T_0^{true}-\sqrt{1-e^2}\Omega_b^{-1}\epsilon_B,\nonumber\\
&& \epsilon_B\equiv \frac{c B}{a_p\sin i}\approx 
1.6\times 10^{-8}\cos\eta.
\label{eq:t_rel_long}
\end{eqnarray}
where the numbers are for the J0737-3039 system.
The degeneracy with Roemer delay implies that to detect aberration 
delay one must look for arrival time variation over longer timescales.
As a result of geodetic precession, both angles $\eta$ and $\zeta$,
and thus $a^{obs},~e^{obs},~T_0^{obs}$ will change with amplitude
$\sim \epsilon_A,\epsilon_B$, on the timescale of the precession period.
In some systems the precession can be rather rapid (with period of 
tens of years) and, if $\epsilon_A,~\epsilon_B$ are not too small, this 
variation can be detected in long-term observations and 
the pulsar spin orientation can be inferred (Stairs \etal 2004). 
Such measurement would be complicated by the fact that variable 
aberration is not the only contribution to the evolution of
$a$ and $e$: they would also evolve because of the gravitational 
wave emission and possible time-dependent Doppler shift of the system,
see DT92. However, these additional contributions should give rise to
linear non-periodic trends in $\dot a$ and $\dot e$ making it 
possible to disentangle their effect in long-term timing observations 
from periodic (nonlinear) behavior of aberration time delay.

In the case of the PSR J0737-3039 system, the precession period of 
pulsar A is $75$~yr, and the relative variation of observed 
eccentricity and semimajor axis on this timescale is 
$3.3\times 10^{-6}\sin\eta$, while the observed epoch of 
periastron varies by $22~\mu$s. The current timing accuracy of J0737-3039A 
(Stairs 2005, private communication) allows determination of 
$e^{obs}$ and $a^{obs}$ with relative accuracy of 
$1.7\times 10^{-5}$ and $1.4\times 10^{-6}$, respectively,
and the measurement of $T_0^{obs}$ has absolute accuracy of $35$~ms. 
Thus, it would in principle be possible to detect variation 
of $a^{obs}$ (but not of $e^{obs}$ or $T_0^{obs}$) caused by 
geodetic precession and longitudinal aberration at a 2$\sigma$ 
level after tens of years of timing observation.

Much better measurement can be made if one monitors the long-term 
drift of timing positions of {\it individual} pulse features instead of 
the whole profile. Then the amplitude of drift\footnote{In this approach 
the orbital parameters should be determined separately for each 
pulse component.} in 
$a^{obs}$, $e^{obs}$, $T_0^{obs}$ caused by geodetic precession 
is augmented by the much more significant latitudinal aberration.
Since the latitudinal delay has the same dependence on the orbital 
elements as the longitudinal one, its effect on variation 
of $a^{obs}$, $e^{obs}$, $T_0^{obs}$ can be included by simply 
changing $A\to A+C$ and $B\to B+D$ in (\ref{eq:ae_rel})
and (\ref{eq:t_rel_long}). As a result, $a^{obs}$ and $e^{obs}$ 
determined from timing of individual pulse features vary on geodetic
precession timescale with relative amplitude 
$4.2\times 10^{-5}(\cos\eta-0.08\sin\eta)$. This should give a 
highly significant detection of geodetic variation of $a^{obs}$ 
(and on a shorter baseline than the longitudinal aberration variation) 
and yield a measurement of $e^{obs}$ variation at a $2\sigma$ level. 
The accompanying variation of $T_0^{obs}$ at the 
level of $0.4$ ms would still escape detection. 

Time delays caused by gravitational light bending do not suffer from 
the degeneracy with Roemer delay 
(or any other timing 
contribution\footnote{The only degeneracy is with the frame-dragging 
delay caused by the companion's spin (Wex \& Kopeikin 1999; 
Rafikov \& Lai 2005) but in binaries consisting of two neutron stars 
the frame-dragging delay is always negligible.}) 
because of their unique time dependence
(see eqs.~[\ref{eq:dtl}], [\ref{eq:delt_L_col}], and 
Figs.~\ref{fig:pos_image}-\ref{fig:neg_image}).
As noted before, in general the rotational lensing delay is 
asymmetric with respect to the conjunction, which may facilitate the 
separation of this delay from the geometric delay and the
(corrected) Shapiro delay (Lai \& Rafikov 2005).
For the parameters adopted in Figure \ref{fig:pos_image}, 
the maximum amplitude of $(\Delta t)_L$ for the positive image 
is $\sim 5~\mus$ (for $i=90.28^\circ$ or the minimum 
$R=4000$~km), while $(\Delta t)_L^{\rm (lat)}$ reaches as high as 
$70~\mus$, both achieved within several seconds of conjunction. 
The latter delay would be detectable if one times the arrival of
individual features of the pulse profile.

Detecting the lensing-related time delays may be a challenging 
task. One factor which limits the accuracy of pulsar timing arises
from the intrinsic variation of the pulse profile from one pulse to
the next. If one times the arrival of a pulse feature with width
$w$, then the timing accuracy after combining $N_p$ pulses is
$\sigma_t\sim w/\sqrt{N_p}$. For PSR J0737-3039A, the pulse
has two peaks, each with width $w\sim 4$~ms. The lensing effect
lasts about $\sim 4$~s, so that in each orbit there are $N_p\sim 200$ pulses 
during the lensing event, and about half of them
\footnote{Lensing effects manifest themselves only very near the 
conjunction (when $R\sim R_E$) where absorption by magnetospheric
plasma of pulsar B is important (Kaspi \etal 2004;
Rafikov \& Goldreich 2004). At the same time, McLaughlin \etal (2004) 
and Lyutikov \& Thompson (2005) have 
demonstrated that even very close to A's superior conjunction 
radio signal of A can still pass through B's magnetosphere almost unattenuated 
at some intervals of B's rotation phase within which at least several 
rotational periods of A can be detected. It is this penetrating 
part of A's radio flux that can be used to search for gravitational 
lensing effects in J0737-3039.}  
may be used for timing purposes. This gives 
$\sigma_t\sim 0.4$~ms for a single passage
through conjunction, somewhat larger than  
the amplitude of $(\Delta t)_L^{(\rm lat)}$ (for $i=90.29^\circ$).
By adding up pulses from many orbits one will obtain better
timing accuracy which may make the detection of $(\Delta t)_L^{(\rm lat)}$ 
possible. This would set a robust constraint on the system's inclination 
angle.

Finally, all rotational effects (both aberration and lensing) 
that we described in application to J0737-3039A should also be relevant for 
J0737-3039B. Since the masses of the two pulsars are quite similar, the 
variation of the direction of emission vector $\Delta {\bf N}$ is 
also similar for both pulsars, but the spin period of pulsar B, $P_B=2.8$~s 
is significantly longer than $P_A=23$ ms. As a result, the 
amplitudes of all spin-dependent delays for pulsar B are $\sim 10^2$ higher 
than for pulsar A. However, the current timing accuracy of pulsar B, 
$2.7$ ms (Lyne \etal 2004), is much poorer than that of pulsar A
and is not expected to improve significantly in the future.
The pulse profile of pulsar B is unstable and subject to variations 
throughout the orbit, presumably due to the influence of 
the intense wind/radiation from pulsar A (Ramachandran \etal 2004).
Moreover, the spin and magnetic geometry of pulsar B (not well constrained 
observationally at present) may not be very favorable for exhibition of large 
latitudinal effects
\footnote{A detailed modeling of the modulation 
of the pulsar A eclipse by pulsar B's rotation yields the following 
parameters for pulsar B: the angle between the spin axis and magnetic axis
$\sim 75^\circ$, the inclination of the spin axis to the orbital angular
momentum axis $\sim 60^\circ$ (Lyutikov \& Thompson 2005).}.

\section{Discussion}
\label{sect:disc}

Pulse profile distortion associated with the latitudinal aberration
delay is the most easily measurable among all rotational 
timing effects in J0737-3039A if the spin-magnetic orientation of this 
pulsar inferred by Demorest \etal (2004) is correct. 
This effect should be detectable with the current timing 
accuracy if the arrival times of individual pulse components  
(rather than the pulse centroid) are monitored.
However, if this effect is not found in the data, we can suggest two
possible reasons. First, it is conceivable that at the present 
time the spin orientation is such that $\cos\eta\ll 1$, 
resulting in a rather small profile distortion by aberration 
(see eq.~[\ref{eq:C_and_D}]). 
This would additionally account 
for the remarkable stability of J0737-3039A 
pulse profile (Manchester \etal 2005). 
However, such a situation would not last since geodetic precession 
of the pulsar spin axis changes $\eta$ on the timescale of years, so 
that several years from now latitudinal aberration should become observable. 
There is also a possibility that 
the spin of pulsar A is almost aligned with the orbital angular momentum
\footnote{The natal kick on pulsar B (e.g. Willems \etal ~2004) would in 
general make $\bs_A$ misalign with the orbital angular momentum axis
except if the kick was directed along or close to the orbital plane.}
which, because of the edge-on orientation of the system, 
would naturally make $\cos\eta\ll 1$. This requires a very 
large opening angle of the emission cone, $\rho\sim 90^\circ$, 
which may be unlikely. 
Second, it is possible that the spin and magnetic orientations
inferred by Demorest \etal (2004) are incorrect, e.g. 
because of the failure of RVM. This would not be very surprising 
given that the polarization data 
are corrupted by quasi-orthogonal modes. Even the interpretation of
the emission as coming from a single magnetic pole can be 
called into question\footnote{This would also affect constraints 
obtained by Manchester \etal (2005) which were derived assuming 
single pole configuration.}. 
If this were the case, the angles
$\alpha$ and $\zeta$ could be different from what we 
used in our estimates, possibly reducing the
magnitude of pulse distortion. Thus, measuring the pulse variation 
in J0737-3039A on the orbital timescale can be used to 
probe the emission and spin geometry of pulsar A.

Pulse shape variations caused by latitudinal shifts
are very promising from the observational point of view: unlike 
the lensing effects, profile dilation due to orbital aberration 
does not require special orientation of orbital plane which makes 
it potentially observable in many binary systems. As far as we are aware, 
no one has attempted to detect pulse profile expansion/contraction 
of the predicted form correlated with the phase of pulsar orbital motion. 
The only possible exception is the study of PSR B1534+12 by 
Stairs \etal (2004) in which pulse shape variations
were detected through fitting a linear combination 
of two standard profiles to all observed profiles. Variation of 
coefficients of this linear combination on short (orbital) and long
(geodetic precession) timescales allowed Stairs \etal to set constraints 
on the pulsar spin geometry. This procedure seems to
be less straightforward than a direct measurement of the variation 
of the phase separation between different pulse components that
we propose to perform, although it has an advantage of yielding  
pulsar spin orientation free from assumptions about the beam geometry.

It may be possible to infer the PSR J0737-3039A
spin orientation from rotational effects alone, without 
resorting to polarimetric observations. 
The pulse profile variations on the orbital timescale and the long-term 
geodetic variations of orbital parameters may be 
enough to measure A's spin orientation with respect 
to its magnetic axis (angle $\alpha$) and with respect to orbital
angular momentum of the binary. Detection of lensing
effects would refine the accuracy of such 
measurement and additionally constrain the inclination of the system, 
and may also help verify the assumption of circular beam shape.

Rotational delays may be measurable in other compact 
binary pulsar systems with significant $v/c$. 
In such systems pulse profile variations on orbital timescales may be
a telltale signature of latitudinal aberration.
The effect is strongest for pulsars in which the polarization angle of 
radio signal does not exhibit significant variation across the pulse 
profile, implying small value of $\tan\chi_0$ and large $\Delta\Phi_0$
(see eq.~[\ref{eq:delphi0}]). 
This could serve as a useful criterion for 
selecting potential targets to detect orbital pulse profile 
variations (geodetic variations will also be enhanced for such 
pulsars by latitudinal effects, see \S \ref{sect:detect}). 
Timing of individual pulse components needed for such detection should 
be facilitated in pulsars with rich pulse profiles exhibiting 
many symmetric sharp features. Using this technique to measure pulsar 
orientation with high significance may require stacking together 
observations obtained in many orbits\footnote{Time 
interval during which such observations are taken must be small compared to
the geodetic precession period since otherwise system's geometry 
would change.}, but, presumably, this would still 
require much shorter baseline than is needed
for detection of timing variations caused by geodetic precession. 

Our results on the lensing effects have been obtained in the
static approximation, which neglects the change in the
orbital configuration of the binary during the time it takes a radio
photon emitted by pulsar A to cross the system. This is
the approach adopted by Damour \& Taylor (1992) and many
other authors since then. Recently Kopeikin \& Sch\"{a}fer (1999)
have gone beyond the conventional static approximation and presented 
ways of including the effect of finite photon propagation 
time on pulsar timing. We plan to investigate the influence of this 
effect on lensing timing signals in the future (Rafikov \& Lai 2005).

\acknowledgements 

We thank Jim Cordes and Ingrid Stairs for useful discussion
and Vicky Kaspi and Rene Breton for making data on pulse 
profile of J0737-3039A available to us. 
We are grateful to an anonymous referee for useful suggestions
which improved this manuscript.
RRR thankfully acknowledges the financial support by the 
W. M. Keck Foundation and NSF via grant PHY-0070928. DL 
thanks CITA and TIARA (Taiwan) for hospitality and is supported 
in part by NSF grant AST 0307252 and NASA grant NAG 5-12034.


\begin{center}
\begin{deluxetable}{ l l l l l l l l l l }
\tablewidth{0pc}
\tablecaption{Time delays for various spin geometries
of J0737-3039A. 
\label{table}}
\tablehead{
\colhead{$\alpha$}&
\colhead{$\zeta$}&
\colhead{$\eta$}&
\colhead{$\tan\chi_0$}&
\colhead{$A$, $\mu$s}&
\colhead{$B$, $\mu$s}&
\colhead{$C$, $\mu$s}&
\colhead{$D$, $\mu$s}&
\colhead{max$|(\Delta t)_L|$, $\mu$s}&
\colhead{max$|(\Delta t)_L^{\rm (lat)}|$, $\mu$s}
}
\startdata
$4^\circ$ & $50^\circ$ & $45^\circ$ & 0.081 & -3.4 & 0.016 & 42.0 & 0.20 & $5.8$ & $72$ \\
$18.7^\circ$ & $78.5^\circ$ & $85^\circ$ & 0.31 & -3.7 & 0.0016 & 1.1 & 0.059 &
$3.1$ & $17$ \\
$5.7^\circ$ & $83.7^\circ$ & $85^\circ$ & 0.091 & -3.7 & 0.0016 & 3.5 & 0.20 &
$3.0$ & $56$ \\
$34.2^\circ$ & $42.3^\circ$ & $85^\circ$ & 0.88 & -5.4 & 0.0023 & 0.54 & 0.030 &
$4.5$ & $8.4$ \\
\enddata
\end{deluxetable}
\end{center}

\end{document}